\documentclass[%
 reprint,aps,prl,superscriptaddress,showpacs
]{revtex4-1}
\usepackage[caption=false]{subfig}
\usepackage{graphicx}
\usepackage{dcolumn}
\usepackage{amsmath}
\usepackage{bm}
\usepackage{tensor}
\usepackage{textcomp}

\begin{document}

\preprint{APS/123-QED}

\title{Performance analysis of $d$-dimensional quantum cryptography under state-dependent diffraction}

\author{Jiapeng Zhao}
\email{jzhao24@ur.rochester.edu}
\affiliation{The Institute of Optics, University of Rochester, Rochester, New York, 14627, USA}
\author{Mohammad Mirhosseini}
\affiliation{California Institute of Technology, Pasadena, California, 91125, USA}
\author{Boris Braverman}
\affiliation{Department of Physics, University of Ottawa, Ottawa, Ontario, K1N 6N5, Canada}
\author{Yiyu Zhou}
\affiliation{The Institute of Optics, University of Rochester, Rochester, New York, 14627, USA}
\author{Seyed Mohammad Hashemi Rafsanjani}
\affiliation{The Institute of Optics, University of Rochester, Rochester, New York, 14627, USA}
\author{Yongxiong Ren}
\affiliation{University of Southern California, Los Angeles, California, 90007, USA}
\author{Nicholas K. Steinhoff}
\affiliation{The Optical Science Company, Anaheim, California, 92806, USA}
\author{Glenn A. Tyler}
\affiliation{The Optical Science Company, Anaheim, California, 92806, USA}
\author{Alan E. Willner}
\affiliation{University of Southern California, Los Angeles, California, 90007, USA}
\author{Robert W. Boyd}
\affiliation{The Institute of Optics, University of Rochester, Rochester, New York, 14627, USA}
\affiliation{Department of Physics, University of Ottawa, Ottawa, Ontario, K1N 6N5, Canada}

\date{\today}

\begin{abstract}
Standard protocols for quantum key distribution (QKD) require that the sender be able to transmit in two or more mutually unbiased bases. Here, we analyze the extent to which the performance of QKD is degraded by diffraction effects that become relevant for long propagation distances and limited sizes of apertures. In such a scenario, different states experience different amounts of diffraction, leading to state-dependent loss and phase acquisition, causing an increased error rate and security loophole at the receiver. To solve this problem, we propose a pre-compensation protocol based on pre-shaping the transverse structure of quantum states. We demonstrate, both theoretically and experimentally, that when performing QKD over a link with known, symbol-dependent loss and phase shift, the performance of QKD will be better if we intentionally increase the loss of certain symbols to make the loss and phase shift of all states same. Our results show that the pre-compensated protocol can significantly reduce the error rate induced by state-dependent diffraction and thereby improve the secure key rate of QKD systems without sacrificing the security.
\end{abstract}
\maketitle


\section{Introduction}
\paragraph{}Quantum key distribution (QKD) is considered to be one of the most promising and practical applications of quantum information science \cite{dixon2010continuous, PhysRevLett.98.010504, sibson2017integrated}. It has been studied both theoretically and experimentally since it was proposed by Bennett and Brassard in 1984 \cite{bennett1984quantum}. In early works, researchers focused mainly on 2-dimensional quantum systems, for example, the polarization states of individual photons \cite{bennett1992experimental}. For the past decade, effort has been dedicated to the investigation of higher-dimensional quantum systems \cite{torres2005twisted,mirhosseini2015high,Sit:17}. The benefits of utilizing higher-dimensional quantum systems for QKD include higher information capacity and enhanced robustness against eavesdropping. \par

Orbital angular momentum (OAM) states are attractive candidates for QKD because they intrinsically span an infinitely large Hilbert space. Beams with an azimuthal phase dependence $\exp(i\ell\theta)$ carry an OAM of $\ell\hbar$ per photon, where $\ell$ is the integer OAM quantum number. After the breakthrough work by Allen \emph{et al.} in 1992 \cite{allen1992orbital}, the properties and applications of OAM have been studied in both classical and quantum regimes \cite{wang2012terabit,mirhosseini2016wigner,gao2017distributed,ren2016experimental,ren2017spatially}.\par

 One characteristic of an OAM state is its $\ell$-dependent diffraction \cite{padgett2015divergence}. Because of the state-dependent diffraction (SDD), OAM states with higher $\ell$ will have larger far-field sizes, and acquire more propagation phase (i.e. the Gouy phase of Laguerre Gaussian (LG) states). Thus, in practical free-space communication links, different OAM states will suffer different amounts of loss for a given collection aperture of finite size \cite{lavery2017free,krenn2016twisted}, leading to $\ell$-dependent detection efficiency. Similar problems occur for states in the complementary angular (ANG) basis, which consist of an equal superposition of OAM states with fixed relative phase between adjacent OAM states \cite{mirhosseini2013rapid,mirhosseini2015high}. Due to the SDD, both the amplitude of each OAM state and the relative phase will be modified. Therefore, the received state will be different from the transmitted state, increasing the error rate at the receiver even in the absence of eavesdropper. The adverse effects of SDD in both OAM and ANG bases result in QKD systems less robust against background noise, measurement errors and eavesdropping. Although OAM-based QKD systems have been demonstrated in both laboratory and outdoor environments \cite{mirhosseini2015high,vallone2014free,Sit:17}, the influence of SDD on QKD systems has not yet been adequately addressed in previous work \cite{wang2018towards}. \par

\paragraph{}Here, we investigate the performance of a $d$-dimensional QKD system under SDD using OAM states as the example. The SDD results in an efficiency mismatch in OAM basis and an increased error rate in ANG basis, which results in a lower secure key rate. These SDD-induced defects are quantitatively studied as a function of the Fresnel product $N_f$ in vacuum, which is defined as $N_f = (\pi/4)D_A D_B/(\lambda z)$ \cite{tyler2011spatial}. $D_A$ and $D_B$ are the diameters of the circular transmitting and receiving apertures respectively, $\lambda$ is the wavelength of the light, and $z$ is the propagation distance. We then propose a pre-compensation protocol to minimize the defects. To validate the approach, we experimentally measure the crosstalk matrices for both the pre-compensated protocol and the original protocol, and then estimate the secure key rates in both cases. We find that for a quantum channel with a small Fresnel number product $N_f$ but high-dimensional encoding space, the pre-compensated protocol can significantly reduce the error rate and provide a greater secure key rate per transmitted photon.\par

\section{Security loophole induced by state-dependent diffraction}
\paragraph{}Because of the finite sizes of transmitter and receiver apertures, higher-order OAM states, which have stronger diffraction, will experience greater loss and acquire more propagation phase. To determine the channel transmission efficiency of a specific OAM state, we define a propagation operator $\hat{F}$ that transfers the OAM eigenstate prepared by Alice $|\ell \rangle_A$ to the state received by Bob $|\ell \rangle_B$ (which is also an OAM eigenstate but has a different radial amplitude distribution) as:
\begin{equation}
  |\ell \rangle_B = \hat{F}|\ell \rangle_A.
\end{equation}
The operator $\hat{F}$ includes the effects of propagation in vacuum and the finite apertures at both transmitter and receiver sides. Note that $\hat{F}$ only results in different amounts of loss, but does not introduce any crosstalk between different OAM $|\ell\rangle$ states. Therefore, this is not a unitary transformation, and if we define the efficiency $\varepsilon_{\ell}$ as $\varepsilon_{\ell} = \langle \ell|\ell \rangle_B/\langle \ell |\ell \rangle_A$, we obtain the following eigenvalue relation \cite{tyler2011spatial}:
\begin{equation}
    \hat{H}|\ell \rangle_A = \varepsilon_{\ell}|\ell \rangle_A,
\end{equation}
where $\hat{H}= \hat{F}^\dagger\hat{F}$ and $\varepsilon_{\ell}$ is the corresponding efficiency of $|\ell \rangle_A$ (See Supplementary material for details).\par

\begin{figure}[h!]
\subfloat[{}]{\includegraphics[width=0.4\textwidth,keepaspectratio]{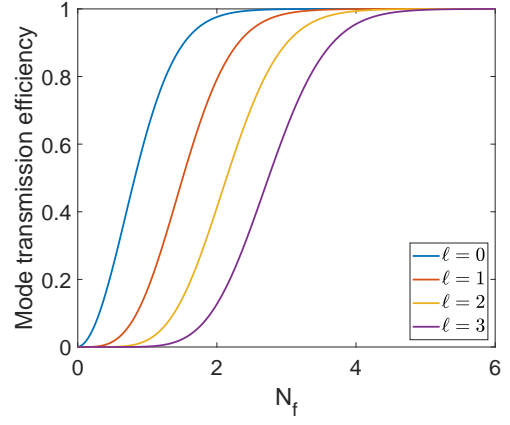}}\\
\subfloat[{}]{\includegraphics[width=0.4\textwidth,keepaspectratio]{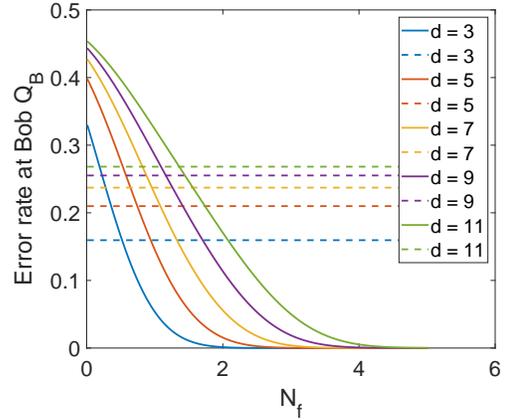}}
\caption{{(a).  State transmission efficiency of different OAM states in a $d=7$ quantum system. (b) the QSER at Bob as a function of Fresnel number product $N_f$. $d$ is the dimension of the Hilbert space. The solid lines show the QSER due to the effects of SDD. The dashed line show the maximum value of the QSER for which a secure channel can be obtained in the limit where the effects from SDD are negligible ($N_f\gg1$). When the QSER goes above the upper bounds, the communication system is not secure and the secure data rate goes to zero. }}
\label{fig:fig1}
\end{figure}
\paragraph{}In OAM-based QKD, the complementary ANG basis is the Fourier conjugate of the OAM basis. The ANG state of index $j$ is defined as \cite{mirhosseini2013rapid,mirhosseini2015high}:
\begin{equation}\label{eq:solve}
    |j\rangle =\frac{1}{\sqrt{d}} \sum_{\ell = -L}^L |\ell\rangle e^{-i 2\pi j\ell /d},
\end{equation}
where $d$ is the dimension of the Hilbert space and $L$ is the maximum OAM quantum number in use, which satisfies the relation: $2L+1 = d$. The ANG basis and OAM basis form two mutually unbiased bases (MUBs), and the use of two or more sets of MUBs guarantees the unconditional security of QKD \cite{RevModPhys.89.015002,RevModPhys.81.1301}. \par

 In practice, as we mentioned above, different OAM states will suffer different amounts of diffraction, as do the OAM components of an ANG state. As shown in Fig.1 (a), for low Fresnel number product ($N_f<1$), there are huge efficiency differences between lower-order and higher-order OAM states. This difference results in a nonuniform probability of detecting the OAM states. The ANG basis for Bob will thus be modified as:
\begin{equation}
\begin{split}
     |j\rangle_B = \frac{1}{\sqrt{\varepsilon_j}} \hat{F}|j\rangle_A &= \sum_{p=0}^{d-1} \sqrt{P_{j,p}}|j+p\rangle_A\\
     &= \sum_{\ell = -L}^L \sqrt{P_\ell} |\ell\rangle_A e^{-i \ell (2\pi j /d-\psi(z))},
\end{split}
\end{equation}
where $1/\sqrt{\varepsilon_j}$ is the normalization constant given by $\varepsilon_j = \sum_{\ell}^{d}\varepsilon_\ell/d$, which describes the transmission efficiency of ANG states. The state $|j\rangle_A$ is the ANG state $j$ prepared by Alice, which has the same form as Eq.(3). The state $|j\rangle_B$ is the ANG state received by Bob after being modified by SDD. The quantity $P_{j,p}$ characterizes the crosstalk between ANG states, and is equal to the probability of finding the ANG state $|j+p\rangle$ prepared by Alice in the ANG state $|j\rangle_B$ received by Bob which has been modified by SDD. The quantity $P_\ell$ is the probability of finding the OAM component $|\ell\rangle$ in the modified ANG state $|j\rangle_B$. The quantities $\sqrt{P_{j,p}}$ and $\sqrt{P_\ell/d}$ are related by a discrete Fourier transform, and one can also show that $P_{j,p}$ is independent of $j$ (see Supplementary material for details). The propagation phase $\psi(z)$ is the phase acquired by each OAM state after propagating a distance $z$. One can notice that the state-dependent loss gives rise to a nonuniform probability distribution of the OAM spectrum, while the state-dependent phase terms introduce extra relative phase between the different OAM components of each ANG state. Both of these effects lead to the crosstalk in ANG basis, which will be further exacerbated in current methods for sorting ANG states \cite{mirhosseini2013efficient}. \par

One direct consequence of the increased crosstalk is an increase in the quantum symbol error rate (QSER) at Bob's side. Here we define the QSER as the probability of detecting a photon in a state other than the launched state \footnote{The QSER is not equal to quantum bit error rate (QBER) because one error symbol can yield more than one bit error in high dimensional QKD. The QBER is equal to QSER only in two dimensional encoding since one error symbol gives one bit error.}: $Q_B = 1-(F_{OAM}+F_{ANG})/2$,
where $F_{OAM}$ and $F_{ANG}$ are the fidelities of the OAM basis and ANG basis respectively, defined as $F_{OAM}=|\tensor*[_A]{\langle \ell|\ell \rangle}{_B}|^2$ and $F_{ANG}=|\tensor*[_A]{\langle j|j \rangle}{_B}|^2$. In our case, assuming there is no eavesdropping, $F_{OAM}$ equals unity while $F_{ANG}$ equals $P_{j}$, which means that only the ANG basis suffers an increased QSER \footnote{The SDD does not change the OAM value upon the propagation. Although each OAM state suffers different amount of loss and acquires different phase, the azimuthal phase vortex of each OAM state is maintained during propagation. Therefore, there is no spread in the OAM spectrum but only loss.}. To quantitatively show how the QSER changes with diffraction, we have numerically calculated the probability distribution of $P_\ell$ for Fresnel number product $N_f$ ranging from 0.01 to 5 under different quantum space dimensions in Fig. 1(a). When $N_f$ is close to 0, only the fundamental Gaussian state (the $\ell = 0$ state) can be transmitted. Therefore, the OAM spectrum at Bob will be very narrow, and the ANG spectrum will become uniform, leading to a complete loss of information. As $N_f$ increases, all OAM states will have equal efficiency near 1, indicating that state-dependent loss is negligible. The QSER as a function of $N_f$ is shown in Fig.1 (b). For a given dimension $d$, small $N_f$ can significantly increase the QSER even if there is no quantum attack. This will lead to a lower information capacity (see Supplementary material for details), and make the system more vulnerable to eavesdropping and quantum cloning since the upper bound for the QSER is fixed for each given dimension $d$ \cite{cerf2002security,sheridan2010security}. Moreover, for a given $N_f$, a higher dimensional system will suffer from more crosstalk introduced by SDD. For instance, in Fig.1 (b), the crosstalk for $d = 11$ is three times larger than the crosstalk for $d = 7$ in a $N_f = 2$ system. In addition to the loss of information, higher error rate means that one needs to sacrifice a greater fraction of the raw key to detect the existence of eavesdroppers, because the legitimate parties cannot distinguish the errors generated by eavesdroppers' attack from other errors in the system.  \par

 The nonuniform efficiencies induced by SDD in the OAM basis introduces a detection efficiency mismatch in Bob's detectors, which can be utilized by Eve to control information received by Bob. The security of QKD in the presence of efficiency mismatch has been both theoretically and experimentally studied \cite{lydersen2010hacking,sajeed2015security,winick2018reliable}. Fortunately, measurement-device-independent QKD protocols have been developed to eliminate the loopholes from side-channels including efficiency mismatch \cite{lo2012measurement,braunstein2012side,pirandola2015high}, and one can implement these protocols to remove this SDD induced security loophole. However, these strategies cannot eliminate the effect of SDD (the increased QSER) in the ANG basis. Therefore, a new protocol that can reduce the effect of SDD in both bases needs to be developed.\par


\section{Waist pre-compensation protocol}

From the discussion above, one can conclude that, the non-uniform efficiency induced by SDD leads to a security loophole in the OAM basis due to state-dependent loss, and an increased error rate in the ANG basis caused by both state-dependent loss and phase. Therefore, to reduce the adverse effects of SDD, a uniform efficiency for all encoding states is desirable, which requires adjusting the efficiencies of low-order states to match the high-order states. \par
Here, we propose a new pre-compensation protocol to mitigate these adverse effects. Alice first selects one set of states that she is going to use for encoding, and measures the efficiency of the state $|\ell_{\rm max}\rangle_A$, where $|\ell_{\rm max}\rangle$ is the largest OAM quantum number to be used. To adjust the efficiencies of all low-order states to match the efficiency of $|\ell_{\rm max}\rangle_A$, she can change the beam radius of each low-order state so that each state has a same divergence angle $\alpha_\ell$: $\alpha_\ell \propto (|\ell|+1)/r_{\rm rms}(0)$, where $r_{\rm rms}(0)$ is the root-mean-square (rms) beam radius defined by Ref.\cite{padgett2015divergence}. That is to say, Alice intentionally increases the loss of the low-order states to reduce the state-dependent loss. We call this set of OAM states uniform-energy-loss (UEL) states. Alice then uses these specially prepared OAM states to construct the corresponding ANG basis $j$. After this, a uniform efficiency has been achieved for both OAM and ANG states, and both the efficiency mismatch and the increased QSER for Bob can be significantly reduced. Since the two bases are orthogonal and mutually unbiased throughout the entire propagation distance, the security analysis of this protocol is identical to the one for the BB84 protocol but with a higher uniform channel loss. We name the approach we have just described \emph{waist pre-compensation} (WPC). \par

\begin{figure}
\subfloat[{}]{\includegraphics[width=.45\textwidth,height=.25\textheight,keepaspectratio]{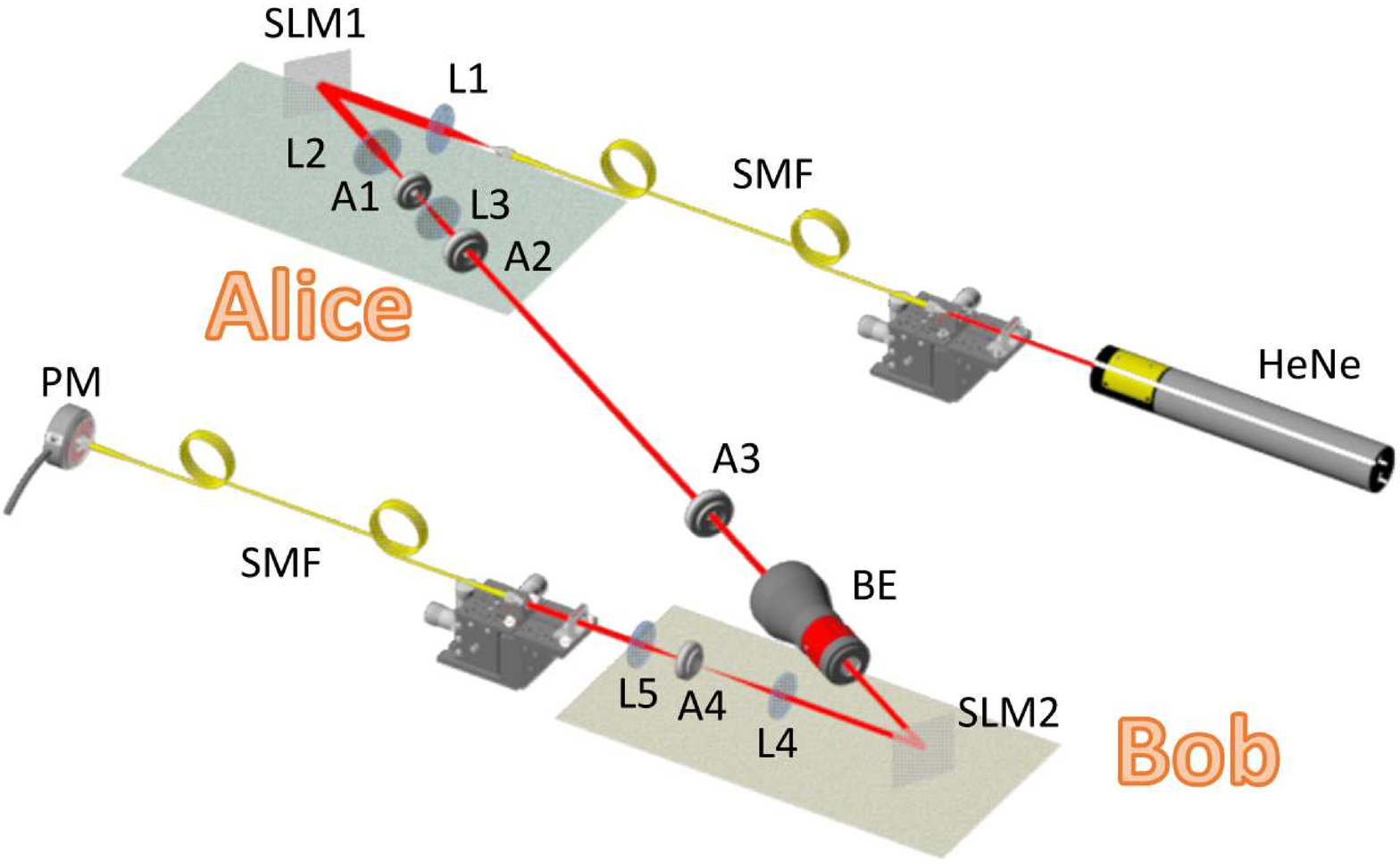}}\\
\subfloat[{}]{\includegraphics[width=.24\textwidth,keepaspectratio]{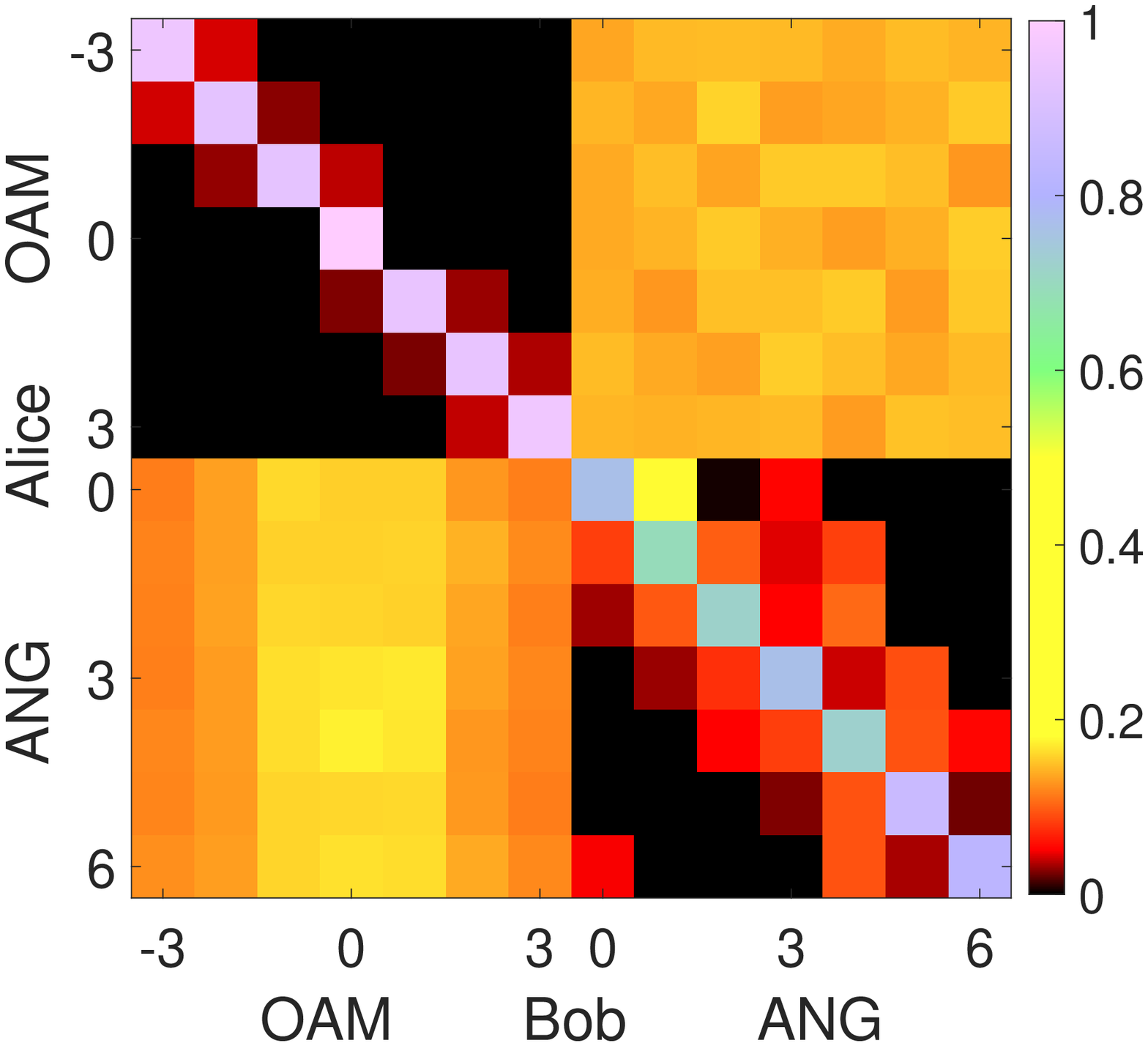}}
\subfloat[{}]{\includegraphics[width=.24\textwidth,keepaspectratio]{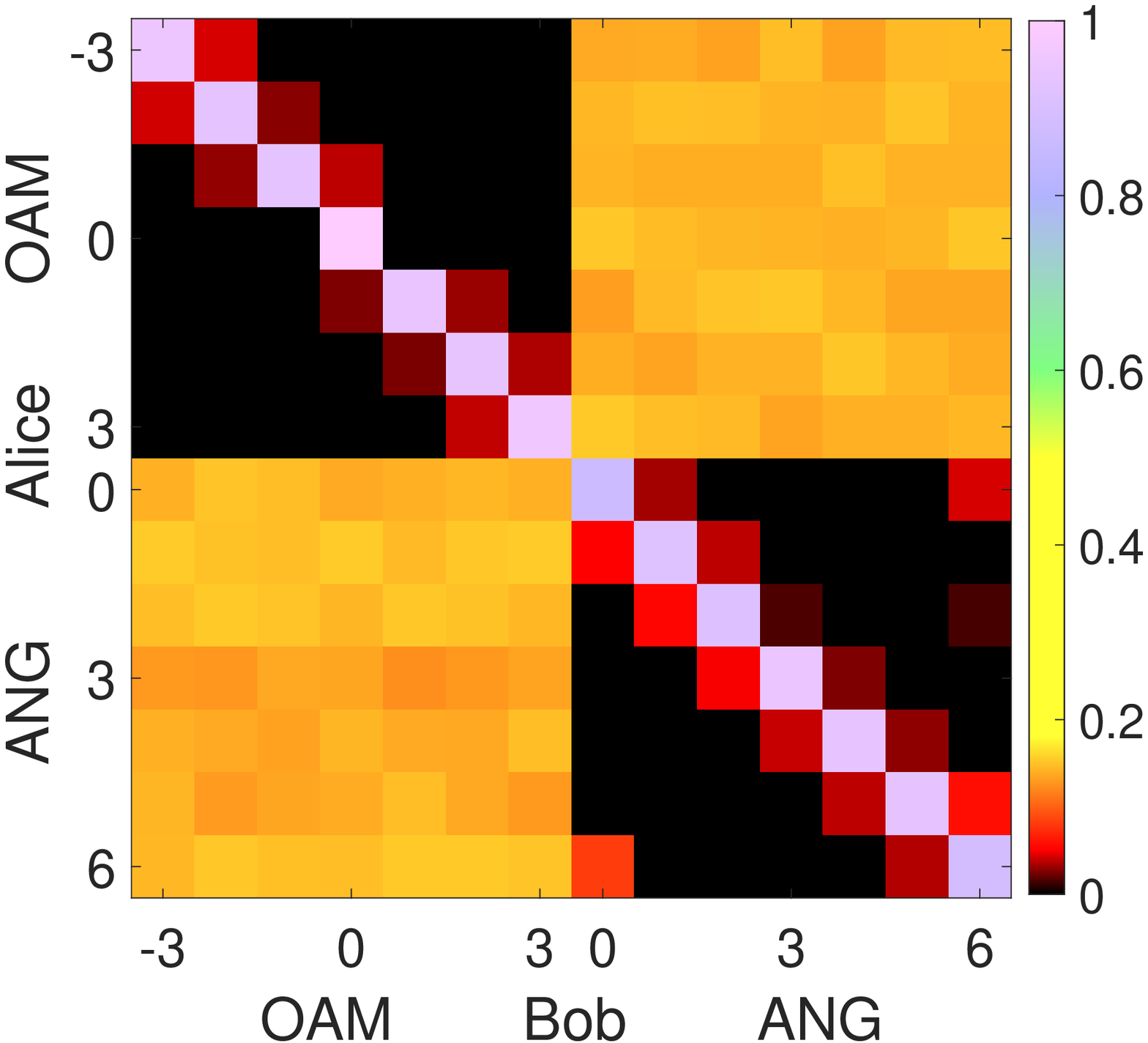}}
\caption{{(a): The experimental setup. L1 to L5 are lenses while SLM denotes the spatial light modulator. A1-A4 are apertures, and BE is beam expander. Z1 represents the propagation distance from transmitter aperture A2 to the receiver aperture A3. (b): the measured probability distribution with no pre-compensation. (c): the experimental result of WPC protocol. We can see that the diagonal elements in figure (c), which represents the fidelity of the states, have less error than in figure (b). The worst fidelity in figure (b) is less than 70$\%$ while the average fidelity is only 85.8$\%$. As the comparison, the worst fidelity in figure (c) is 86.8$\%$ and the average fidelity is 93.3$\%$. Therefore, the crosstalk in figure (c) is less than the crosstalk in figure (b).}}
\label{fig:fig1}
\end{figure}



\section{Experimental results}
\paragraph{}To implement our protocol in a laboratory setting, we measure the crosstalk matrix for a Fresnel number product $N_f = 3.96$ and dimension $d=7$. A HeNe laser is coupled into a single-mode fiber (SMF) to generate a single spatial mode at 633 nm. The first spatial light modulator (SLM1), together with lenses 2 (L2, focal length 0.75 m) and 3 (L3, focal length 0.5 m), are used to generate the desired input states $|\ell\rangle_A$ and $|j\rangle_A$ \cite{mirhosseini2013rapid}. Aperture 1 (A1) is used to select the first diffraction order. The distance (Z1 = 3.12 m) between transmitter's aperture (the diameter of A2 is 3.07 mm) and receiver's aperture (the diameter of A3 is 3.25 mm) constitute the link with Fresnel number product $N_f =3.96$. Both A2 and A3 are implemented by round apertures written onto SLM1 and SLM2 respectively. The second SLM scans the OAM and ANG spectra, and projects the desired state onto the fundamental Gaussian state, which can be coupled into the second SMF. The details of the projective measurement are included in the Supplement. A power meter (PM) is used to measure the transmitted intensity at the end.


\paragraph{}To quantitatively show the benefits of the WPC protocol, we measure the conditional probability of finding each state received by Bob for each state transmitted by Alice, and display the results in a crosstalk matrix (Fig. 2 (c)). One can see that the crosstalk in the ANG basis is very small, in particular when compared with the crosstalk of no compensation protocol, which is shown in Fig. 2 (b). The average QSER measured in the case of no compensation is $14.2\%$ while the average QSER with WPC is $6.7\%$. The mutual information with WPC protocol equals to 2.56 bits per photon, an improvement over 2.22 bits per photon in the case of no compensation. From the QSER above, we can then find the secure key density using the following equation based on two MUBs protocols \cite{sheridan2010security, bradler2016finite}:
\begin{equation}
  r = \log_2d+2[Q\log_2\frac{Q}{d-1}+(1-Q)\log_2(1-Q)],
\end{equation}
where $Q$ represents QSER at Bob's side. The secure key density $r$ is then found to be 1.76 bits per photon with WPC protocol, a significant improvement from 0.89 bits per photon in the no compensation case. \par

\section{Discussion and conclusion}
Although the WPC protocol will ensure the robustness of the quantum system and provide a higher information encoded per photon, it will lower the overall efficiency and may result in a lower secure key rate because of a lower average transmission probability \cite{mirhosseini2015high}. However, for practical quantum encoding systems with higher dimensionality and lower Fresnel number products, the crosstalk introduced by SDD can be much larger than in the ideal case. In such circumstances, the external errors from either modal dispersion (from turbulence, optical fiber, etc.) or imperfect mode sorting can be severe, which makes it even more necessary to implement the pre-compensation protocol for a better QKD performance. \par

The simulated comparison of the secure key rate per transmitted photon between WPC protocol and no compensation protocol with different error rates from external errors are shown in Fig. 3 as a function of Fresnel number product $N_f$. When $N_f$ is small, we see that WPC protocol can significantly improve the performance of high-dimensional QKD systems in realistic links, especially in the presence of external errors. Even for our in-laboratory measurements, which have low external errors, the secure key rate is increased from 0.86 to 1.63 bits per transmitted photon when WPC is implemented (see Supplementary material for details). Therefore, in realistic QKD systems, intentionally sacrificing some efficiency for low-order states to get a lower but uniform efficiency can significantly benefit the system. \par

\begin{figure}
\subfloat[{}]{\includegraphics[width=0.4\textwidth,keepaspectratio]{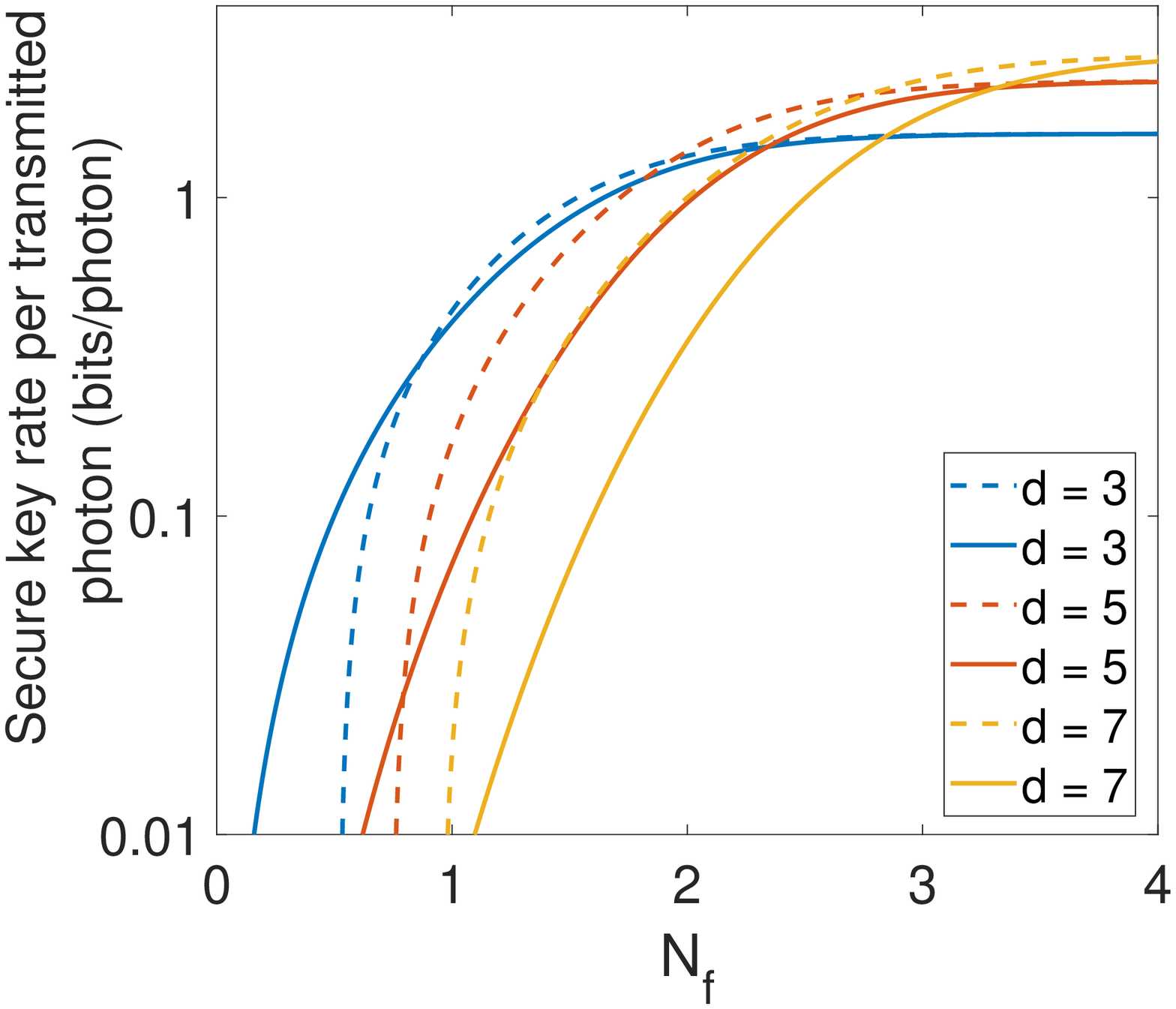}}\\
\subfloat[{}]{\includegraphics[width=0.4\textwidth,keepaspectratio]{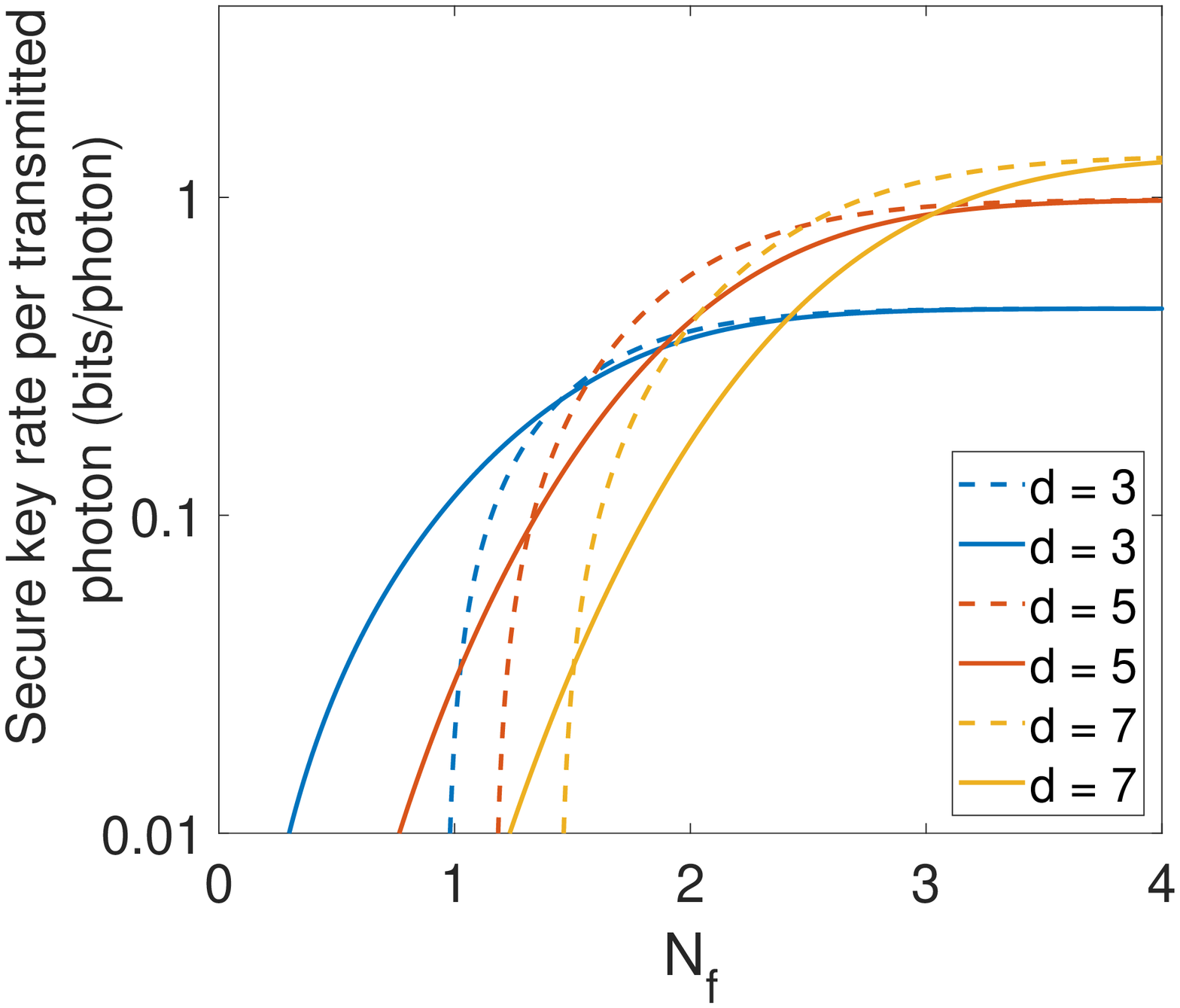}}

\caption{{ (a) and (b): Simulated secure key rate per transmitted photon as a function of Fresnel product $N_f$ with 0$\%$ and 10$\%$ external errors respectively. The solid lines represent the secure key rate using WPC protocol while the dashed lines represent the secure key rate with no compensation.}}
\label{fig:fig2}
\end{figure}

Another concern regarding the WPC protocol is how practical it will be in a realistic QKD scenario. As we discussed above, in most cases, the WPC protocol is superior to no compensation protocols only when $N_f$ is limited. In realistic scenarios, most current free space QKD systems have Fresnel number products less than 1, especially the satellite-to-ground system ($N_f = 0.23$) \cite{schmitt2007experimental, liao2017satellite}. Thus the WPC protocol could be useful in optimizing the performance of future global high-dimensional QKD systems. Furthermore, implementing WPC protocol is simple: one only needs to take the $N_f$ of the system into consideration, and employ the optimal set of beam waists, which requires no extra apparatus.\par

We assumed OAM encoding and circular apertures in the discussion given above. However, SDD is expected to be a problem for any type of spatial-mode encoding, and for a given system, we can always find a set of eigenstates with uniform transmission efficiency. Therefore, our new protocol is generic for realistic high dimensional quantum encoding scenarios utilizing spatial degrees of freedom \cite{xie2016experimental,zhou2017sorting,zhou2018high}.\par

In summary, we have analyzed the performance of a high-dimensional QKD system based on OAM encoding in the presence of SDD. In practical free-space quantum links with finite aperture sizes and long transmission distance, SDD can introduce a high error rate and security loopholes, which can significantly reduce the information capacity of the quantum link and its robustness against quantum attacks. To overcome this threat, we propose the use of WPC based on the use of UEL states, which have a uniform loss for all encoding states. We implemented this approach experimentally and showed that it can appreciably reduce the QSER and improve the secure key rate per transmitted photon. Since the two bases in the WPC protocol are orthogonal and mutually unbiased, the security of this new approach is the same as the conventional BB84 protocol. Therefore, by intentionally increasing the loss for certain states to get a uniform efficiency, we can significantly reduce the adverse effects induced by SDD, and improve the secure key rate in QKD systems. Considering that in the near future, high-dimensional QKD systems will be a promising platform for increasing the channel information capacity of free-space communication systems, our WPC protocol will aid in improving the performance of these systems, and increase their robustness to eavesdropping. \par

\paragraph{}We acknowledge helpful discussions with Boshen Gao, Cong Liu and Kai Pang. This work is supported by U.S. Office of Naval Research. R.W.B. acknowledges support from Canada Research Chairs Program and the National Science and Engineering Research Council of Canada.


\appendix{}
\section{Derivation of the eigenvalue Eq.(2) in the main paper}
\paragraph{}We employ the same notation as in the main paper. The $|\ell \rangle_A$ represents the state prepared by Alice while $|\ell \rangle_B$ denotes the state received by Bob. The propagation operator $\hat{F}$ includes both the effects of diffraction and limited aperture size at the receiver. That is, this operator is not unitary and includes the loss of the link. Therefore, if we define the power sent out by Alice as $P_A = \tensor*[_A]{\langle \ell|\ell \rangle}{_A}$ and the power received by Bob as $P_B = \tensor*[_B]{\langle \ell|\ell \rangle}{_B}$, we can write the efficiency $\varepsilon_{\ell} = P_B/P_A = \langle \ell|\ell \rangle_B/\langle \ell |\ell \rangle_A$. We can then rewrite this equation in the following form:
\begin{equation}
\begin{split}
 \tensor*[_B]{\langle \ell|\ell \rangle}{_B} & = \tensor*[_A]{\langle \ell|\hat{F}^\dagger\hat{F}|\ell \rangle}{_A}\\
& =\tensor*[_A]{\langle \ell|\varepsilon_{\ell}|\ell \rangle}{_A}.
  \end{split}
\end{equation}
If we define a new operator $\hat{H} = \hat{F}^\dagger\hat{F}$, we can get the following relation:
\begin{equation}
  P_B = \tensor*[_A]{\langle \ell|\hat{H}|\ell \rangle}{_A}.
\end{equation}
Therefore, to find the optimal field maximizing the efficiency, we need to maximize the $P_B$ for a given $P_A$. Then, we use a Lagrange multiplier by introducing an additional scalar variable $\varepsilon_{\ell}$, and rewrite this optimization problem as:
\begin{equation}
   \widetilde{P} = \tensor*[_A]{\langle \ell|\hat{H}|\ell \rangle}{_A}-\varepsilon_{\ell}[\tensor*[_A]{\langle \ell|\ell \rangle}{_A}-P_A].
\end{equation}
By differentiating $\widetilde{P}$ with respect to $\tensor*[_A]{\langle \ell|}{}$, we have:
\begin{equation}
  \frac{\partial \widetilde{P}}{\partial\tensor*[_A]{\langle \ell|}{}} =  \hat{H}|\ell \rangle_A-\varepsilon_{\ell}|\ell \rangle_A.
\end{equation}
By optimizing the $\widetilde{P}$, the Eqn. (4) should be 0. Therefore, we have the following eigenfunction:
\begin{equation}
  \hat{H}|\ell \rangle_A = \varepsilon_{\ell}|\ell \rangle_A.
\end{equation}
Hence, the resulting eigenvalue $\varepsilon_{\ell}$ is the efficiency of the eigenstate $|\ell \rangle_A$ which maximizes the transmission efficiency of a given link.\par
\section{Calculation of mutual information $I_{AB}$ in the absence of eavesdropper}
\paragraph{}First we prove that the $\sqrt{P_{j,p}}$ and $\sqrt{P_\ell/d}$ are related by a quantum Fourier transform, and $P_{j,p}$ is independent of ANG state index $j$. From Eq.(4) in the main paper, we can find the following inner products:
\begin{equation}
\begin{split}
  \tensor*[_A]{\langle j+p|j \rangle}{_B} &= \sqrt{P_{j,p}} \\ &= \tensor*[_A]{\langle j+p|}{}\sum_{\ell = -L}^L \sqrt{P_\ell} |\ell\rangle_A e^{-i \ell (2\pi j /d-\psi(z))}.
  \end{split}
\end{equation}
Therefore, we can find the following relation:
\begin{equation}
\begin{split}
  \sqrt{P_{j,p}} &= \sum_{m = -L}^L \sqrt{\frac{1}{d}} \tensor*[_A]{\langle m|}{}e^{i 2\pi (j+p) m /d} \\ &\times\sum_{\ell = -L}^L \sqrt{P_\ell} |\ell\rangle_A e^{-i \ell (2\pi j /d-\psi(z))}\\
  & = \sum_{\ell = -L}^L \sqrt{\frac{P_\ell}{d}}e^{i 2\pi (j+p) \ell /d}e^{-i 2\pi j\ell /d}e^{i \ell \psi(z)}\\
  & = \sum_{\ell = -L}^L \sqrt{\frac{P_\ell}{d}}e^{i  \ell 2\pi (p /d+\frac{\psi(z)}{2\pi})}.
  \end{split}
\end{equation}
Therefore we can find the $P_{j,p}$ is independent of $j$ and equals the Fourier transform of $\sqrt{P_\ell/d}$.\par
We then show how to get the Eq.(5) in the paper. The probability of Alice sending out each symbol is still equal, but due to the state-dependent loss, the probabilities of finding each symbol at Bob's side are different. Therefore, as what we discussed in our paper, for the photons which are registered by both parties, we have the following probabilities:
\begin{equation}
\begin{split}
  P(OAM_{\ell,B}) = P_\ell, \\P(ANG_{j,B}) = 1/d, \\ P(OAM_{\ell,A}) = P_\ell, \\ P(ANG_{j,A}) = 1/d.
  \end{split}
\end{equation}
$P(OAM_{\ell,B}$ represents the probability that Bob receives a photon in $|\ell\rangle$ state, while $P(OAM_{\ell,A}) = P_\ell$ represents the probability that Alice sends out a photon in $|\ell\rangle$ state. This is because those events that Alice sends out one symbol but Bob receives nothing have been discarded.\par
The definition of mutual information is:
\begin{equation}
    I_{AB}=\sum_{a\in A}\sum_{b\in B}p(a,b)\log_2\frac{p(a,b)}{p(a)p(b)},
\end{equation}
where $p(a,b)$ is the joint probability. The relation between joint probability and conditional probability is: $p(a,b) = p(a)p(b|a)$. In our case, even if Alice is sending out each symbol with equal probability, the photon statistics at Bob's side are not uniformly distributed because of the state-dependent loss. Therefore, we have the following probability relations:
\begin{equation}
  p(\ell_A,i_B) = P_\ell\delta_{i \ell}, p(j_A,k_B) = P_{j,p}/d.
\end{equation}

Therefore, considering Alice randomly chooses her basis, the mutual information between Alice and Bob $I_{AB}$ equals:
\begin{equation}
  I_{AB} = \frac{1}{2}I_{AB,OAM}+\frac{1}{2}I_{AB,ANG},
\end{equation}
where $I_{AB,OAM}$ represents the mutual information using OAM basis while $I_{AB,ANG}$ is the mutual information using ANG basis. The final form of $I_{AB}$ can be found as:
\begin{equation}
     I_{AB} = \frac{1}{2}\sum_{p}P_{j,p}\log_2P_{j,p}d-\frac{1}{2d}\sum_\ell \log_2P_\ell.
\end{equation}
As shown in Fig. 4, it is not difficult to verify that $I_{AB}$ is smaller than the ideal value $\log_2d$. When $N_f$ is near zero, the information encoded is almost lost, while in the high $N_f$ region, the information capacity gets close to the ideal value. Another interesting result is that the information carried by the two bases is not equal, and that the information encoded in the OAM basis is always larger than that carried in the ANG basis because of the absence of crosstalk in the OAM basis. \par
\begin{figure}
\subfloat[{}]{\includegraphics[width=0.4\textwidth,keepaspectratio]{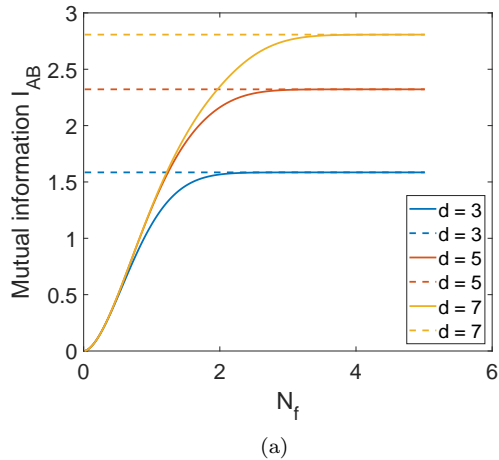}}
\caption{(a): The mutual information between Alice and Bob as a function of $N_f$. The solid lines represent the mutual information with SDD while the dashed lines indicate the mutual information $\log_2(d)$ in the limit where SDD can be ignored ($N_f \gg 1$).}

\label{S1}
\end{figure}

\begin{figure}
\subfloat[{}]{\includegraphics[width=.45\textwidth,keepaspectratio]{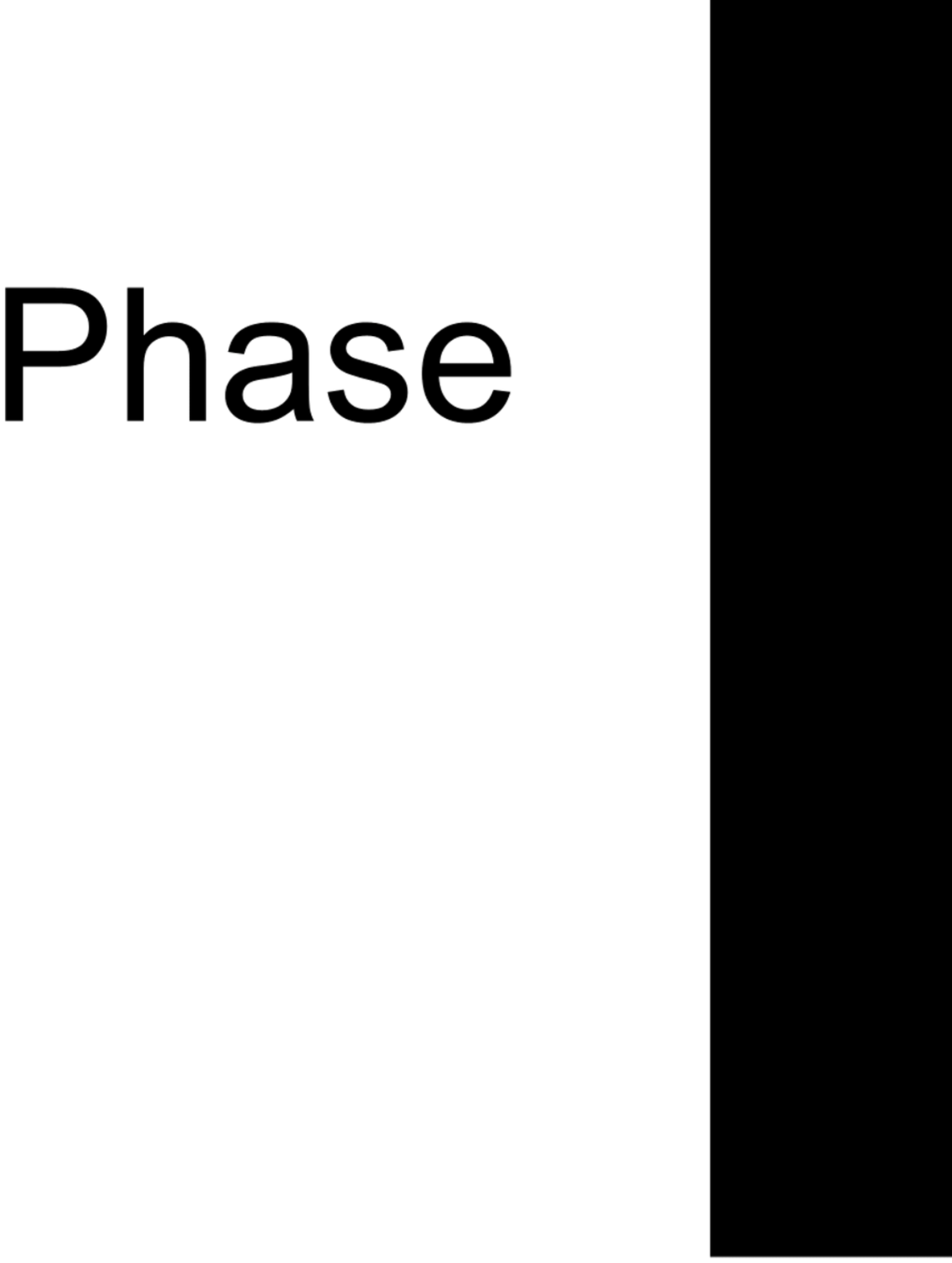}}\\
\subfloat[{}]{\includegraphics[width=.45\textwidth,keepaspectratio]{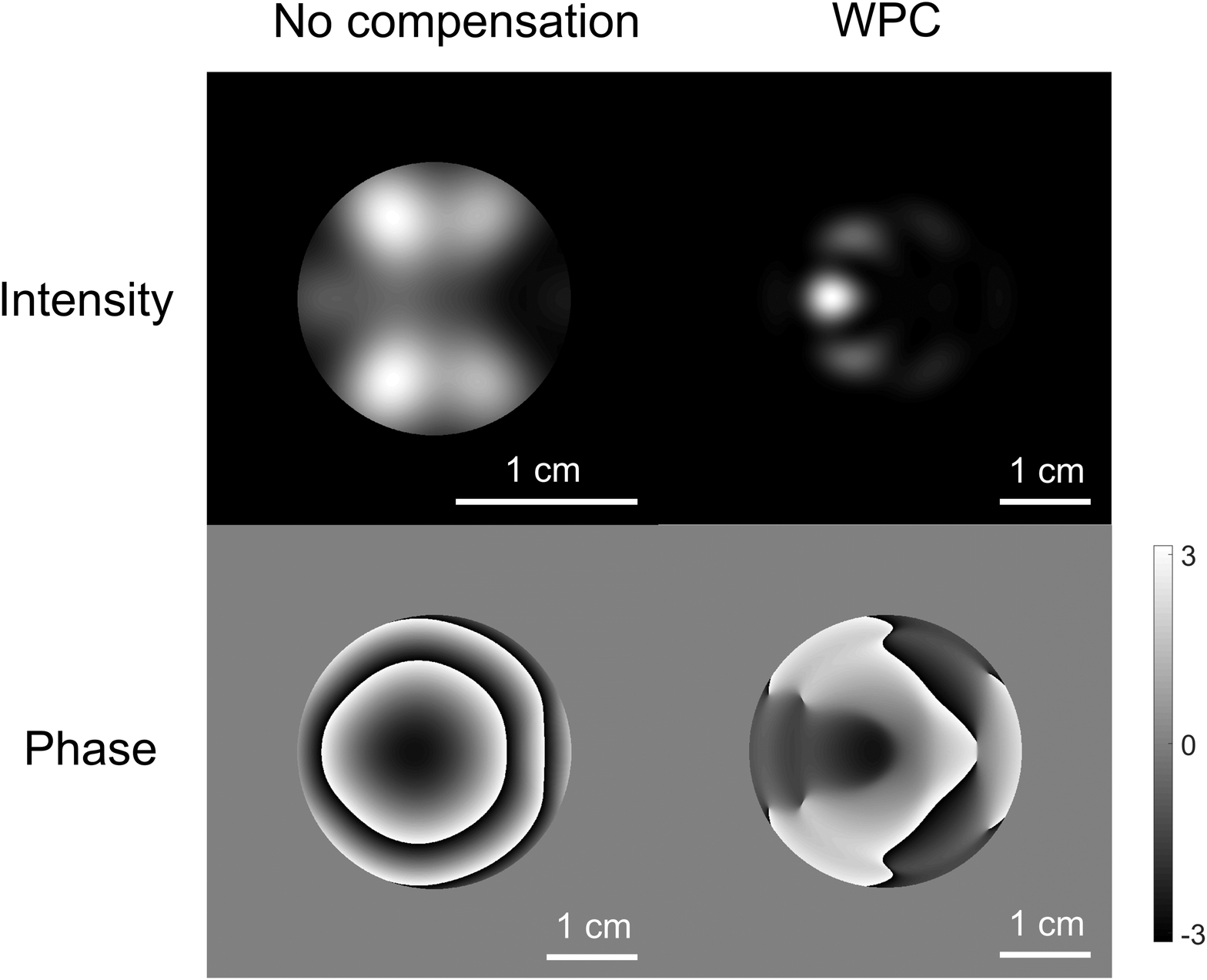}}\\
\subfloat[{}]{\includegraphics[width=.25\textwidth,keepaspectratio]{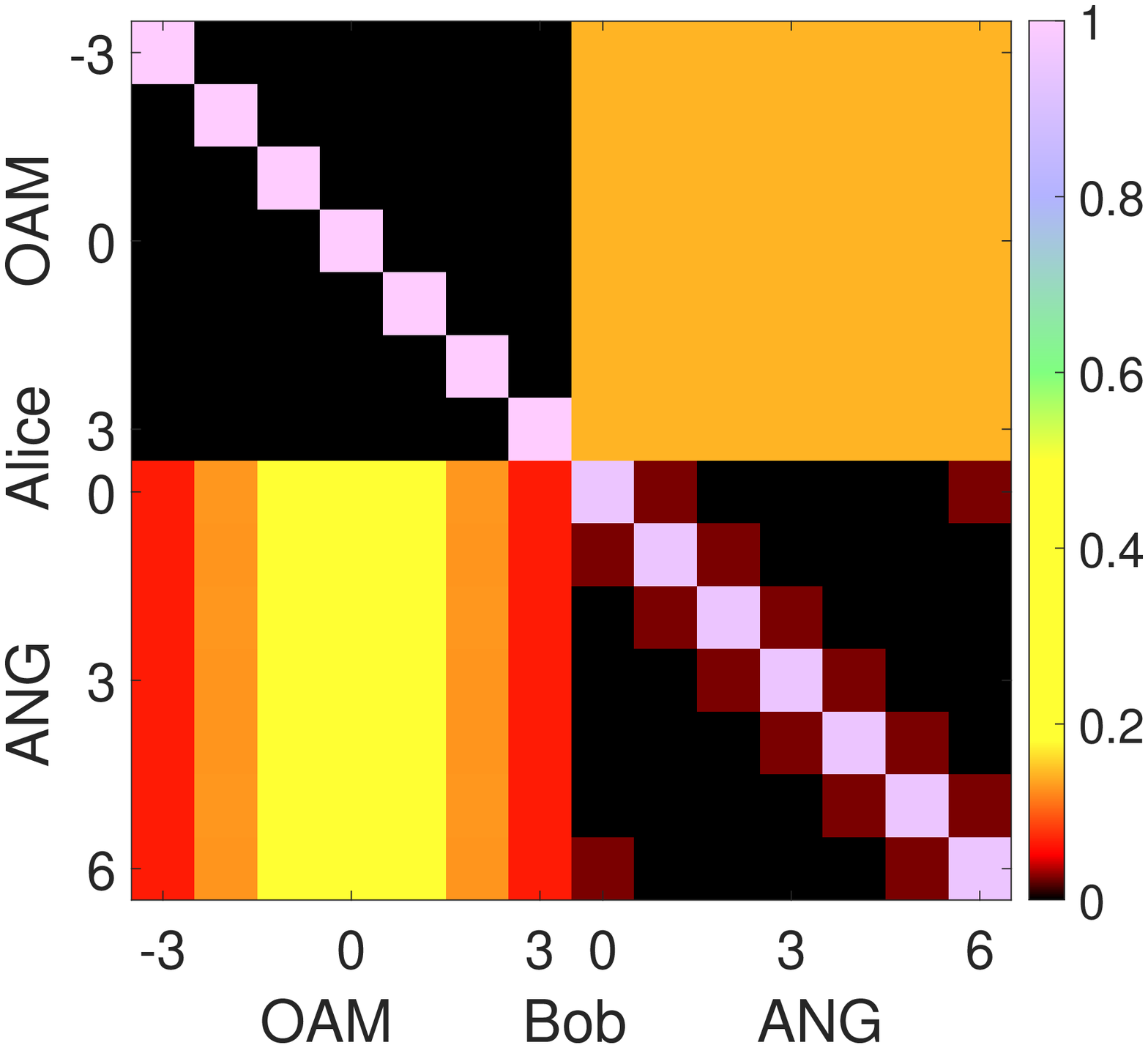}}
\subfloat[{}]{\includegraphics[width=.25\textwidth,keepaspectratio]{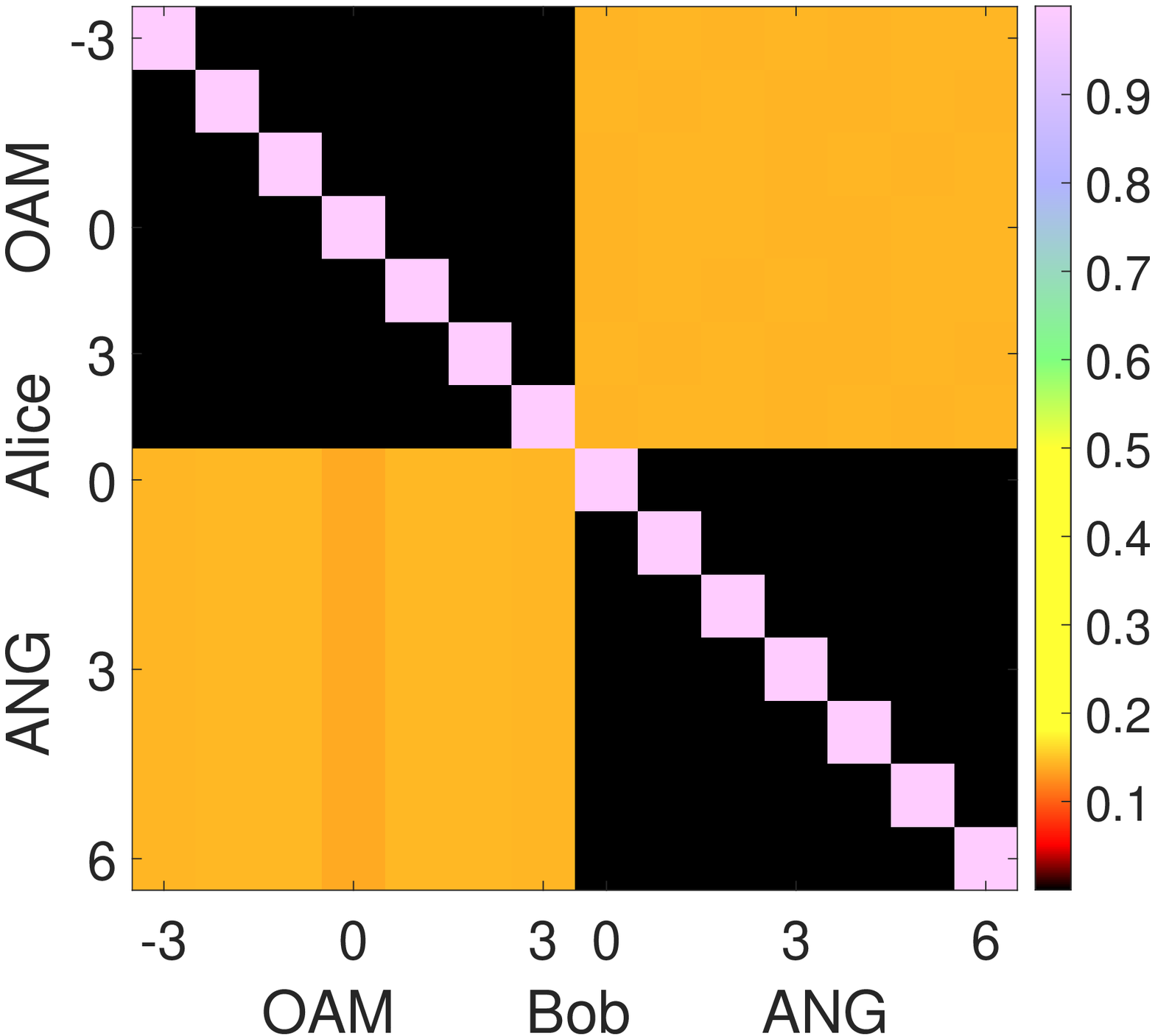}}
\caption{{(a) and (b): Simulation results of transmitted and received $|j=0\rangle$ states respectively, both with WPC and without compensation. (c) and (d): probability distributions of finding each OAM component in the received ANG state $|j=0\rangle$, in the no-compensation protocol (c) and WPC protocol (d). The simulated link has a Fresnel number product equal to 3.96.}}
\label{S2}
\end{figure}

\section{Simulation results of WPC protocol}
\paragraph{}Fig. 5 shows simulation results comparing the WPC protocol and transmission without the use of compensation, which we will refer to as the no-compensation protocol. The Fig. 5 (a) shows the simulated intensity and phase distributions of the ANG state $|j=0\rangle$ prepared by Alice in the no-compensation and WPC protocols, while the Fig. 5 (b) shows the corresponding results for the received ANG state $|j=0\rangle$. One can notice that both the intensity and phase profiles for the two protocols are very different at Alice's and Bob's sides. Diffraction distorts the intensity distribution of the received state in the no-compensation case; after propagating through the link, the one single main lobe on Alice's side, which indicates the angular position and the value of $j$, becomes two main lobes on Bob's side. In contrast, the intensity profile in the WPC case remains similar even after diffraction. The simulated crosstalk matrices of no compensation protocol and WPC protocol are shown in Fig. 5 (c) and (d), respectively. It is obvious that in the no compensation case, the SDD gives a nonuniform probability distribution when we measure the ANG states in the OAM spectrum, and nonzero off-diagonal elements in the ANG spectrum (the fidelity of ANG states shown in Fig. 5 (c) is 95.1$\%$ since the Fresnel number product is chosen to be 3.96). However, with the WPC protocol, an almost uniform probability distribution can be found when ANG states are measured in OAM basis, and there are no nonzero off-diagonal elements in the ANG basis (the fidelity of ANG states in Fig. 5 (d) is 99.99$\%$). Therefore, the simulation results shows the ability of WPC protocol to reduce the adverse effects of SDD.\par

\section{Prepared and received states}
Fig. 6 shows the images of experimentally realized ANG states after transmitting and receiving apertures for $d = 7, \ell_{max} = 3$. The top row gives the ANG states with ANG quantum number $j$ from 0 to 3 prepared by the transmitter. All the states in the top row are prepared with no compensation while the figures in the bottom row are the comparison between no compensation and WPC. Fig. 6 (e) and (f) are prepared and received ANG states $|j = 5\rangle$ with no compensation, and Fig. 6 (g) and (h) are prepared and received states in the WPC case. After diffracting in the link, the spatial profile of ANG state $|j = 5\rangle$ in the no compensation case changes greatly, such that it is intractable to identify the angular position of the main lobe of the state. However, in the WPC case, the received ANG state has a similar spatial profile as the launched state.\par
\begin{figure}
\subfloat[{}]{\includegraphics[width=0.1\textwidth,keepaspectratio]{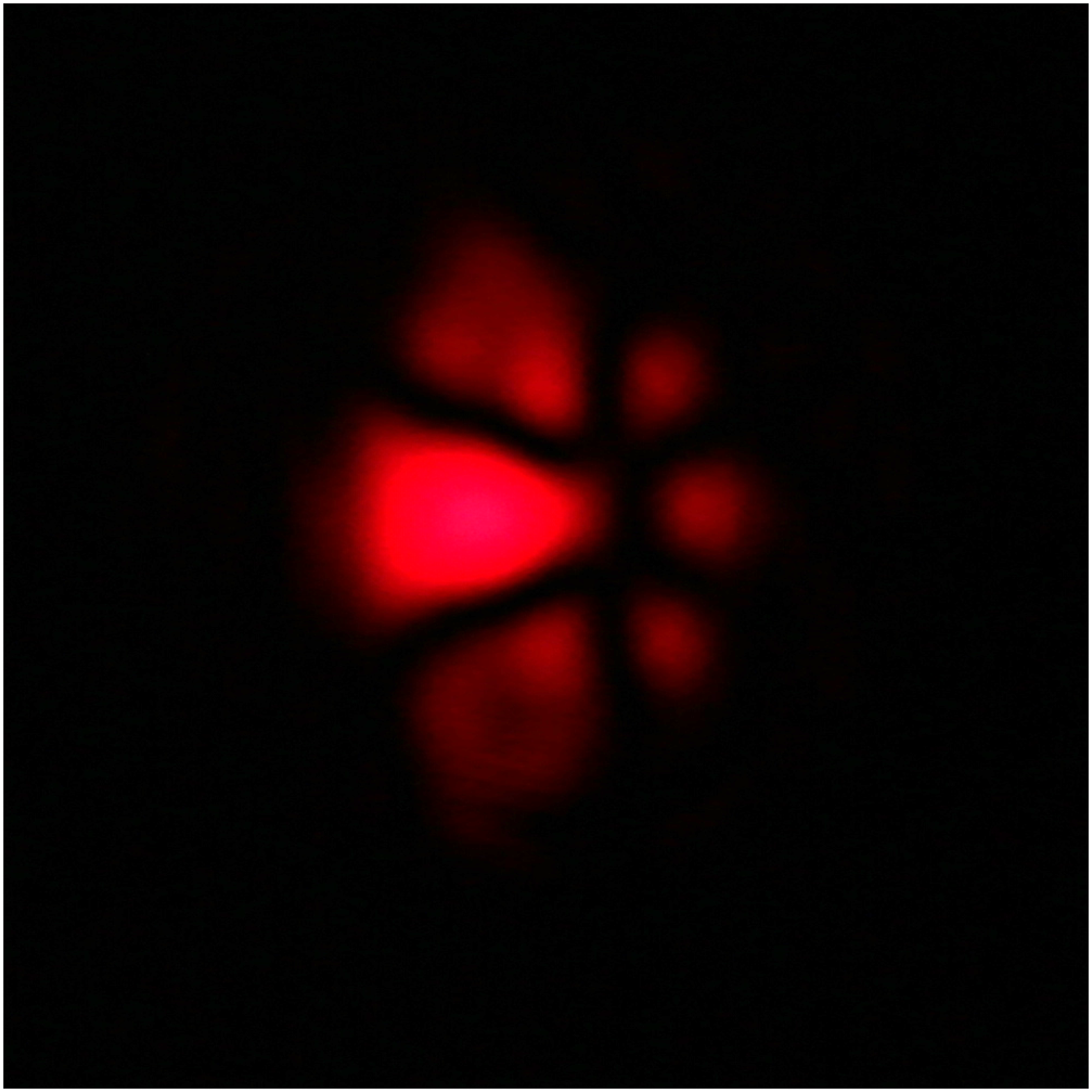}}
\subfloat[{}]{\includegraphics[width=0.1\textwidth,keepaspectratio]{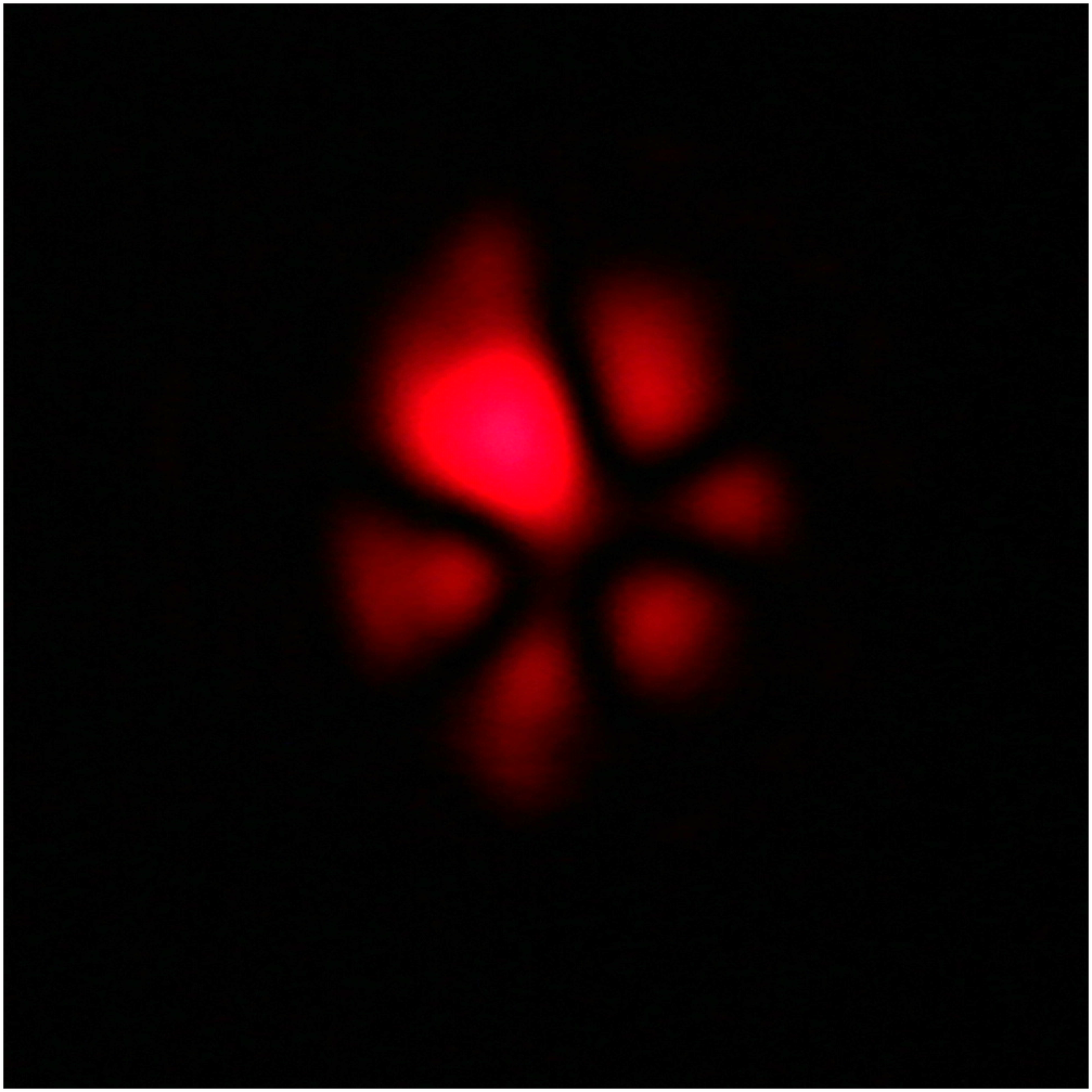}}
\subfloat[{}]{\includegraphics[width=0.1\textwidth,keepaspectratio]{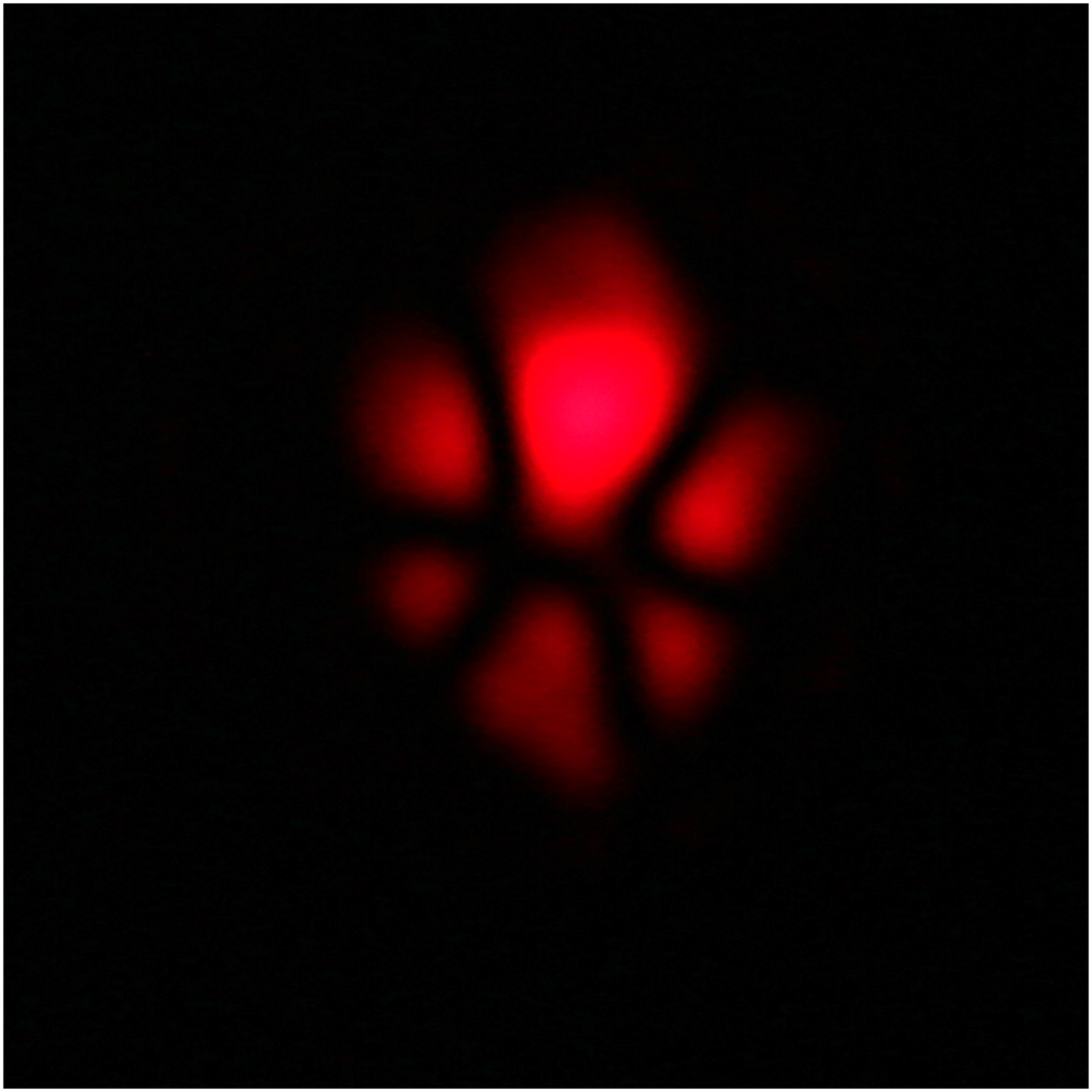}}
\subfloat[{}]{\includegraphics[width=0.1\textwidth,keepaspectratio]{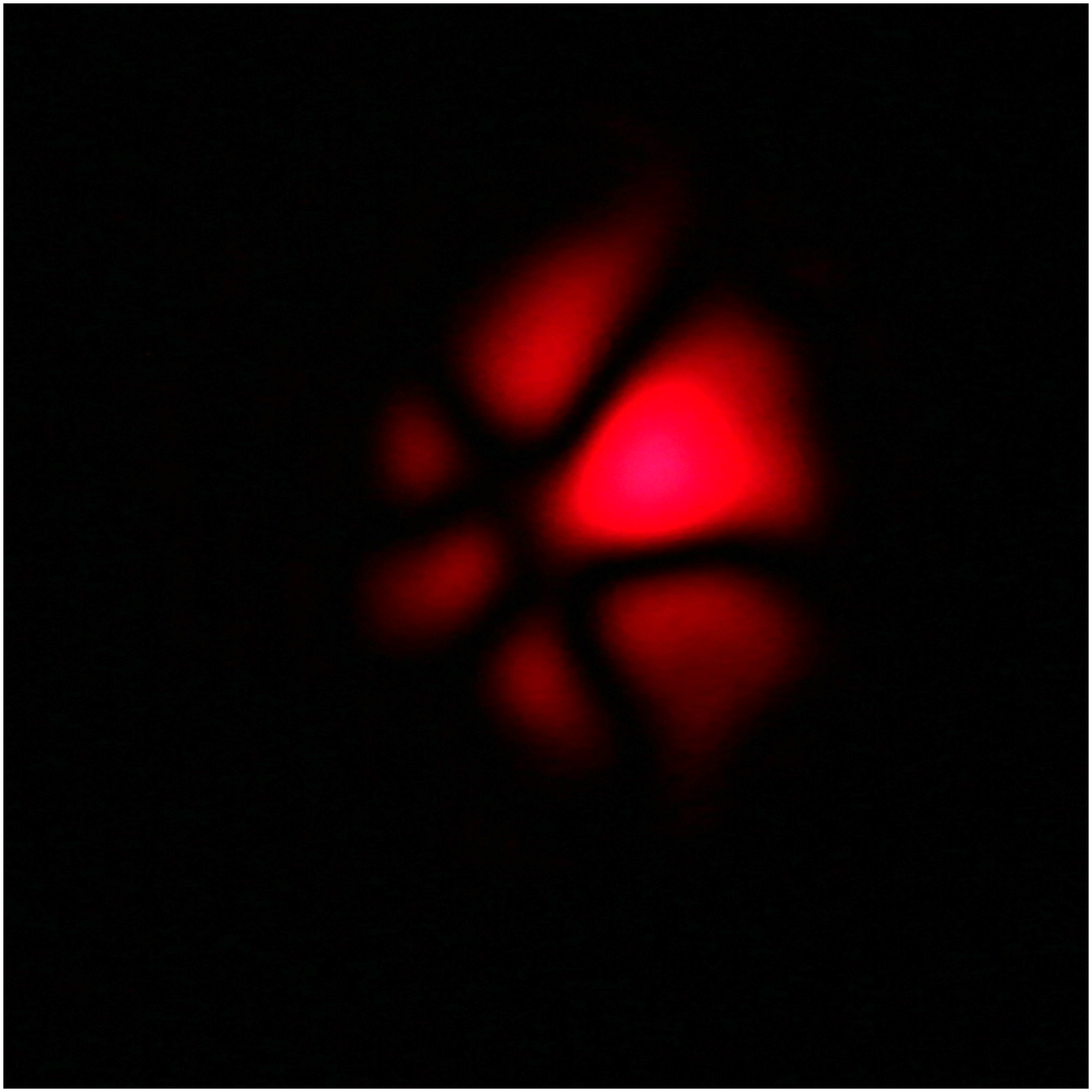}}\\
\subfloat[{}]{\includegraphics[width=0.1\textwidth,keepaspectratio]{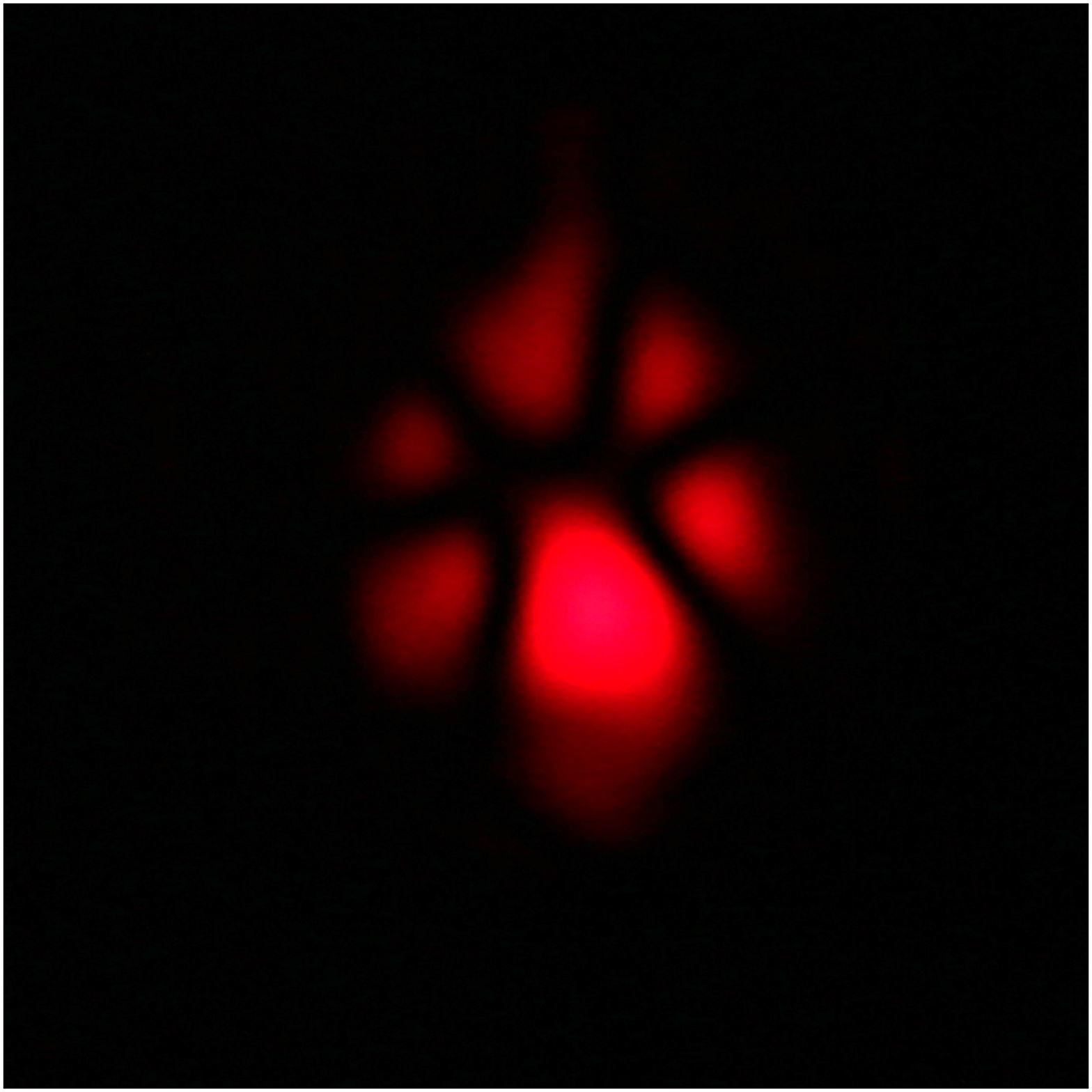}}
\subfloat[{}]{\includegraphics[width=0.1\textwidth,keepaspectratio]{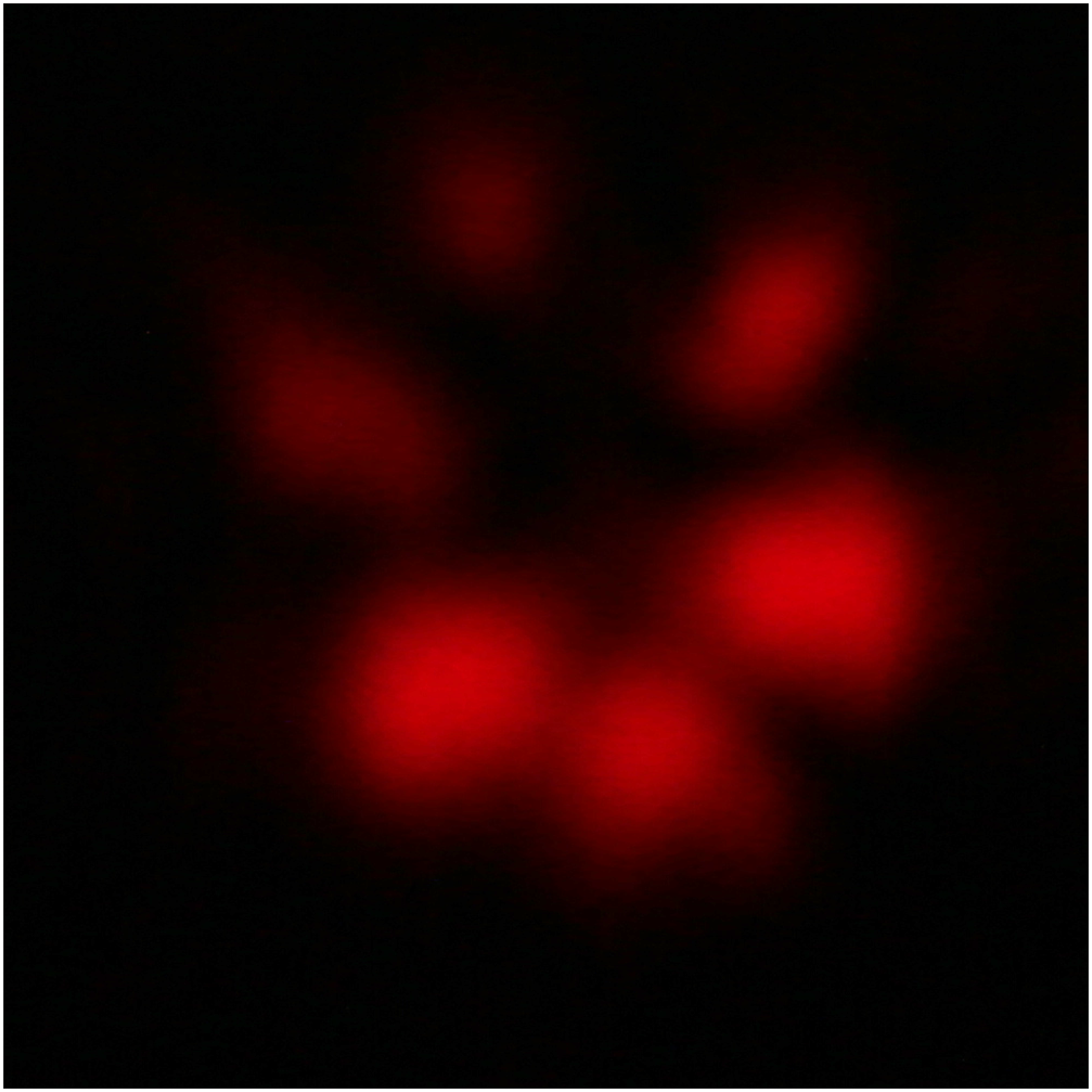}}
\subfloat[{}]{\includegraphics[width=0.1\textwidth,keepaspectratio]{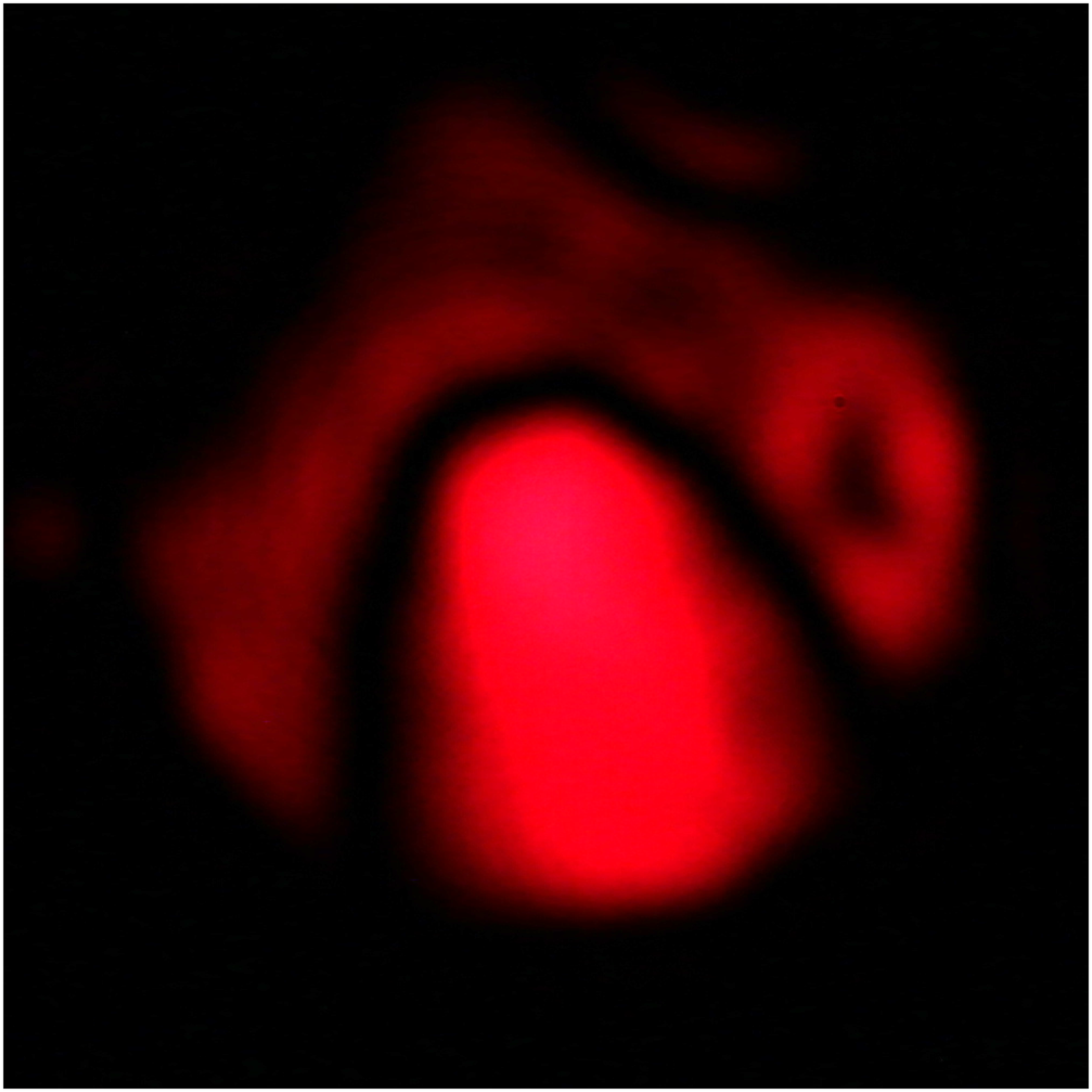}}
\subfloat[{}]{\includegraphics[width=0.1\textwidth,keepaspectratio]{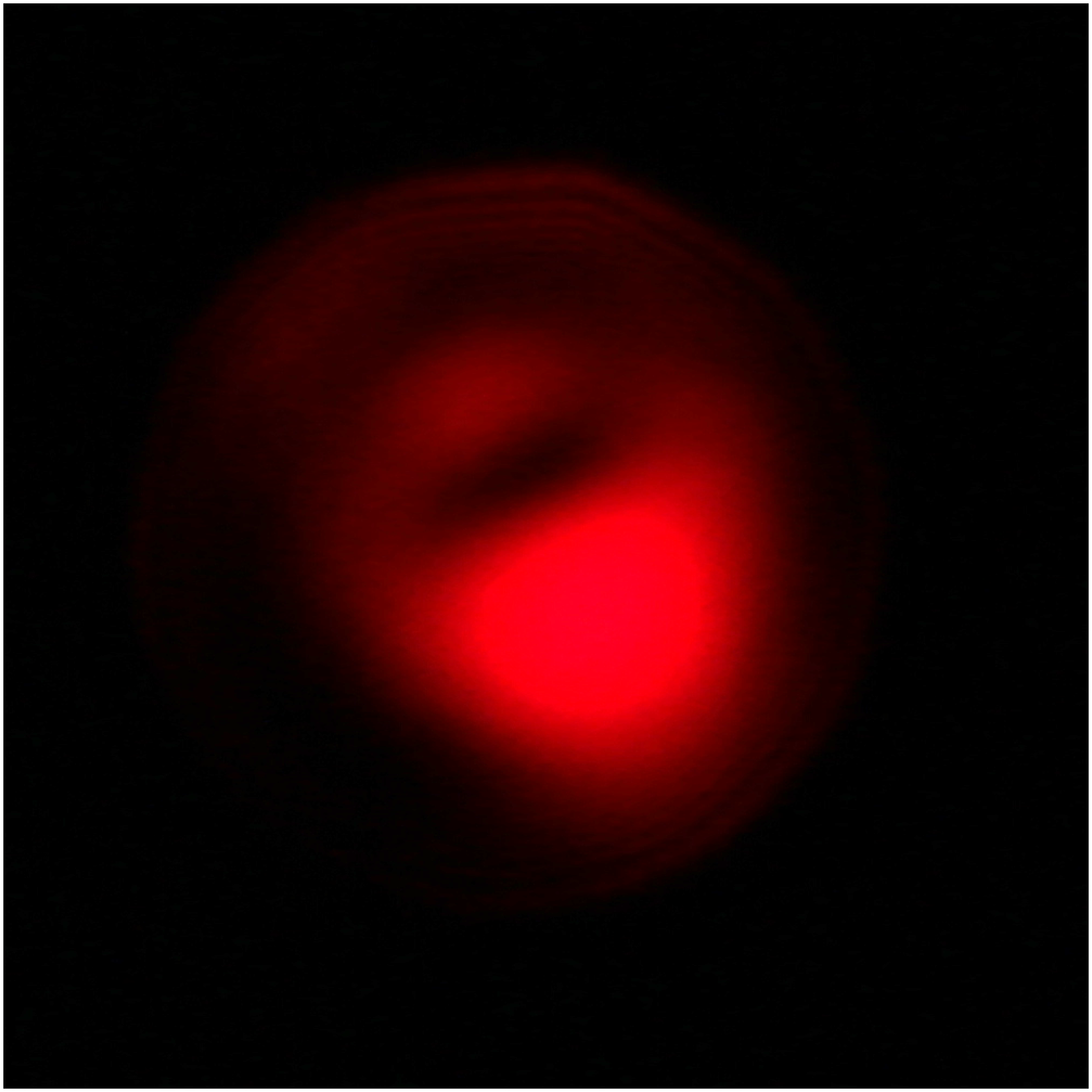}}

\caption{ (a) to (d): Four ANG states generated experimentally with no compensation with $N_f = 3.96$ at transmitter's side. The ANG quantum number of these states is $j = 0, 1, 2$ and $3$ respectively. (e) to (f): the prepared and received ANG states $|j = 5\rangle$ in the no compensation protocol. (g) to (h): the prepared and received states in the WPC protocol. All images are taken under identical acquisition parameters. Note that in panel (f) the dominant lobe of (e) has disappeared, but that it is retained in (h) through the use of pre-compensation. }
\label{S3}
\end{figure}

\section{Projective measurement}
Here, we explain how we experimentally realize the projective measurement in OAM and ANG bases. For the OAM states, as the first step, we use SLM2 to apply diffraction gratings with the same OAM quantum number as the incident beam onto the SLM2. In this case, we couple the Gaussian states which are selected from the first negative diffraction order from SLM2 into the SMF. Since the negative first diffraction order adds the opposite phase we added onto the SLM2, in cases where the OAM quantum number in the incident beam match the OAM value on the SLM2, the beam in the first negative order should be a Gaussian. Therefore, we can record the coupling efficiencies of each incident OAM state by switch the OAM quantum number in the diffraction grating. The single mode coupling efficiency for fundamental Gaussian state is about 40\%. All these calibrations are done with an infinitely large collection aperture. Then, to do the projective measurement of the incident beam in the OAM basis, we sequentially implement the diffraction gratings with different OAM quantum number through the use of SLM2, and then record the powers coupled into SMF. These powers are divided by the corresponding coupling efficiencies of each OAM state to get the exact power of each OAM component in the incident state before coupling. To get the probability distribution we show in the main paper, one needs to normalize the measured power of each incident OAM state. In theory, if the incident beam is in an OAM state, there will be no crosstalk in OAM basis in both conventional protocol and WPC. This is because the grating on the second SLM only modulates the phase of the incident beam but not the amplitude. Therefore, the projective measurement can be described by the following equation:
\begin{equation}
\begin{split}
  &P(\ell_i,\ell_m) = \\ &\frac{\int_{0}^{R}\int_{0}^{2\pi}|A_i(r)exp(i\ell_i\theta)exp(-i\ell_m\theta)|^2 r drd\theta}{\sum_{\ell_m = -\ell_{max}}^{\ell_{max}}\int_{0}^{R}\int_{0}^{2\pi}|A_i(r)exp(i\ell_i\theta)exp(-i\ell_m\theta)r|^2 drd\theta},
\end{split}
\end{equation}
where $P(\ell_i,\ell_m)$ is probability of finding the OAM $\ell_m$ component in the incident beam which has an OAM equal to $\ell_i$. $A_i(r)$ is the radial field distribution of the incident beam. Since the integral over azimuthal degree of freedom gives a Kronecker delta, Eqn. (18) will finally reduce to $P(\ell_i,\ell_m) = \delta_{\ell_i,\ell_m}$. Therefore, the radial field distribution of the incident beam has no influence on the crosstalk in OAM basis when the incident beam is in an OAM state.
For the ANG states, the measurement we did is not a complete projective measurement since ANG states have both amplitude and phase information, but one single SLM can only manipulate one of them. However, we find out that if we use the same method as what we use in the OAM basis, we will only get light coupled into SMF when the ANG quantum added on the SLM2 matches the ANG quantum number of the incident beam in the no SDD case. Even though this method provides a very low coupling efficiency (around 10\%), we can still scan the ANG space and get the crosstalk matrix. However, when the SDD is taken into consideration, one can still find some coupling in the SMF when the ANG quantum numbers mismatch so that we have errors in the ANG basis (i.e. the errors induced by SDD only). This gives the off diagonal terms in the crosstalk matrices.\par

\section{Secure key rate calculation based on experimental data}
The equation (5) in the main paper gives the secure key density of per photon, while the product of the secure key density and the transmission efficiency of the states yields the secure key density per transmitted photon. The secure key rate is then simply the product of the secure key rate per transmitted photon and the photon rate. From the experimental data, the measured efficiency for the $\ell = 3$ state is $ 92.4\%$ which is the transmission efficiency of WPC using UEL states. The average efficiency of all 7 states is $97.1\%$, which is the transmission efficiency of the conventional protocol. Therefore, the secure key density with WPC can be calculated as 1.63 bits per transmitted photon, as compared to 0.86 bits per transmitted photon with no compensation. We can see that a uniform efficiency distribution for all spatial modes, even though it is low, can provide an improved key rate over maximum transmission efficiency. Note that there is a discrepancy in the secure key rate comparison for a $N_f = 4$ system between simulation and lab data (Fig. 5). In theory, the WPC is advantageous only when $N_f$ is small, since for systems with $d=7$ and $N_f=4$, the QSER in the ANG basis is very small (less than 1$\%$). However, in the lab, due to the imperfect measurements, the QSER in the ANG basis is much larger than our prediction (average QSER is $23.6\%$ without compensation, and $8.57\%$ with compensation). The improvement in the QSER in the ANG basis leads to the better performance in WPC protocol, which can improve realistic QKD systems where the QSER in the ANG basis is always larger than in OAM basis \cite{mirhosseini2013efficient, mirhosseini2015high}.
\end{document}